# Growth of α-β phase W thin films over steep Al topography in a confocal sputtering machine

J. M. Kreikebaum[a], B. Cabrera, J. J. Yen

Stanford University, Department of Physics, 382 Via Pueblo, Stanford, CA 94305

P. L. Brink, M. Cherry, A. Tomada

SLAC National Accelerator Laboratory, 2575 Sand Hill Road, Menlo Park, CA 94025

B. A. Young

Santa Clara University, Department of Physics, 500 El Camino Real, Santa Clara, CA 95053

a) Electronic mail: jmkreike@stanford.edu

We report on thin-film processing improvements in the fabrication of superconducting quasiparticle-trap-assisted electrothermal-feedback transition-edge sensors (QETs) used in the design of Cryogenic Dark Matter Search (CDMS) detectors. The work was performed as part of a detector upgrade project that included optimization of a new confocal sputtering system and development of etch recipes compatible with patterning 40 nm-thick, $\alpha-\beta$ mixed-phase W films deposited on 300-600 nm-thick, patterned Al. We found that our standard exothermic Al wet etch recipes provided inadequate W/Al interfaces and led to poor device performance. We developed a modified Al wet-etch recipe that effectively mitigates geometrical step-coverage limitations while maintaining our existing device design. Data presented here include SEM and FIB images of films and device interfaces obtained with the new Al etch method. We also introduce a method for quantitatively measuring the energy collection efficiency through these interfaces.

## I. Introduction

The use of superconducting sensors for a wide range of physics applications has grown dramatically in the past two decades[1-6]. For the Cryogenic Dark Matter Search (CDMS), thousands of interleaved ionization and phonon collection electrodes are photolithographically patterned on both sides of kilogram-scale, high-purity germanium substrates. Detector performance is well understood and dozens of full size detectors have been deployed in physics experiments operated at ~50 mK[7-10]. During a recent fabrication facility upgrade at Stanford, the Balzers barrel-style sputtering system with 5" x 10" targets used for many years to deposit W, Al, and poly-Si for this work was replaced with a sputtering machine of confocal design using 4" diameter targets. Guided by previous rounds of process refinements on the original Balzers system at Stanford/SLAC[11] and during the commissioning of the Segi system at Texas A&M University fabrication facility[12], a series of test devices were fabricated to evaluate physics performance metrics of the films produced using the new machine. Measurements on as-deposited films of Al, W, and poly-Si included film-stress, room temperature electrical resistivity, thin-film superconducting $T_c$, and resistivity ratio, $R_{300K}/R_{4.2K}$. Deposited film thickness uniformity was better than 5% for W and better than 3% for Al and poly-Si over the 10" diameter, four-substrate carrier.

More advanced diagnostic measurements were made on fully patterned W/Al quasiparticle-trap-assisted electrothermal-feedback transition-edge sensor (QET) test devices. Energy transport within and between the Al and W sputtered films was measured at ~ 35 mK in a dilution refrigerator using a collimated $^{55}Fe-NaCl$ x-ray fluorescence source mounted immediately above the QET[13]. This work showed that previously established Al etch recipes used for CDMS detector (and QET) production were not optimal for the films deposited in the new machine. In particular, although film uniformity, resistivity, stress, and superconducting $T_c$ met specification values, the chamber conditions required to deposit these films in our required device geometries (40 nm-thick W sputtered over 300 nm-thick, patterned Al) yielded inadequate W-on-Al sidewall coverage for devices beyond the central ~ 4" diameter region of the substrate carrier. Below, we report on the subsequent development of a new Al wet-etch recipe that consistently provides smooth, high-quality sidewalls that couple well to 40 nm-thick, superconducting W films sputtered over steep Al topography.

## II. Fabrication Processes
### A. QET Test Device Fabrication

Fabrication of CDMS and corresponding QET test devices requires two thin-film deposition sequences (Si-Al-W trilayer; single W film) interspersed with standard micron-scale photolithography[11]. For the QET devices described in this paper, standard 100 mm-diameter, <100> silicon test wafers were diffusion-cleaned using traditional methods ($H_2O:H_2O_2:NH_4OH$, $HF$, $H_2O:H_2O_2:HCl$) before being loaded into a four-wafer substrate carrier with 10" diameter. The wafer carrier was inserted via load-lock into a custom AJA International ATC-2200 confocal DC/RF magnetron sputtering system configured with four, 4"diameter targets in a sputter-up geometry. The chamber and substrate carrier were pre-conditioned *in situ* by sputtering ~0.1 µm Al on all accessible surfaces prior to substrate loading. Depositions were done in 99.999% Ar gas after reaching a chamber base pressure of 8.0E-8 torr.

Clean wafers were lightly RF-etched *in situ* for 15 minutes at 100 W substrate power in 3 mT Ar. A 40 nm-thick layer of 99.999% pure ($>1\Omega-cm$) Si was deposited at 5.5 mT Ar and 400 W target power (480 sec. with shutter open). The as-deposited poly-Si film stress was typically -100 MPa, as determined by comparing wafer curvature before and after depositions on test wafers using a Flexus 2320 stress gauge. The Si deposition was followed immediately by a 300 nm-thick deposition of 99.9998% pure Al performed at 2 mT Ar using 700 W target power (1007 sec.). Lastly, a 99.96% pure, 30 nm-thick W layer was deposited at 3 mT Ar and 400 W target power (226 sec.) with -150 V DC bias on the substrate carrier. The W layer inhibits oxidation of the Al film beneath it to ensure good Al/W film connectivity in the finished devices. The wafers were then removed from the sputtering chamber via load-lock.

Using S1813 photoresist and an EVG620 aligner for patterning, the devices were etched in 30% $H_2O_2$ at 25 º C for 300 sec. to remove exposed portions of the W film, followed by an Al etch that differed in detail from run to run in a controlled way (this study). As discussed more below, the best Al etch recipe for kg scale devices was found to consist of a 300 sec. dip, with agitation, in KMG 80:3:15 NP at 25º C, followed by a series of three, 10 sec. dips in the same 80:3:15 NP bath interspersed with 20 sec. DI water quenching dips (and no dry cycle). The three short follow-up etch/quench cycles were found to be critically important: they smoothed the Al film sidewalls exposed during the initial Al etch. The wafers were then subjected to an additional 850 sec. in 30% $H_2O_2$ (i.e. W etch) to compensate for undercutting that occurred during the Al etch.

After resist stripping, the wafers were reloaded into the AJA sputtering system via load-lock and the chamber was again evacuated to 8.0E-8 torr. The patterned wafers were RF-etched for 15 min in 3 mT Ar at 100 W before a final 40 nm-thick W layer was deposited at 3 mT Ar and 400 W target power (300 sec.). The second W layer was then patterned and etched for 675 sec. in 30% $H_2O_2$. Dies were cut and the quality of the 40 nm-thick W film growth over the much thicker Al layer (300 nm or 600 nm depending on sample set) was inspected with a Hitachi S-4800 SEM. For completeness, some samples were also inspected using FIB imaging techniques. After visual confirmation that the W was continuous and uniform, the devices were tested for physics performance by J. J. Yen, *et. al.* with an $^{55}Fe-NaCl$ fluorescence source ($Cl\ K_\alpha$ at 2.62 keV) in a Kelvinox-15 dilution refrigerator[13].

## *B. Al Etchants*

Three commercially available premixed Al etchants were tested during this study: Cyantek Al-11 (later decommissioned from our shared processing lab), KMG 16:1:1:2 NP, and KMG 80:3:15 NP. Their properties are summarized in Table 1. The phosphoric acid contained in the etchants is responsible for dissolving the few nm of native oxide that grows on the Al film surface and the nitric acid oxidizes the bulk Al which is then removed by the phosphoric acid[14,15]. Acetic acid is used for wetting and buffering[15]. The Al wet-etching process is highly exothermic and this is evident by the increase in bath temperature during etches. Our observations agree with Williams, *et. al.*[14] that agitation aids etch-rate uniformity across the wafer.

As expected, the Al etch rate for our devices was strongly dependent on temperature and was roughly 6 times faster at 40º C compared to 25º C for the etchants tested. Al wet etches are generally isotropic so undercutting of the photoresist mask is fairly common. It is in this etched "pocket" that localized heating occurs, accelerating the Al etchant, and causing further undercutting of the mask[15]. Figure 1 shows this effect in one of our devices where we show the SEM of a device made with Cyantek Al-11 without compensation for the undercut. A room temperature Al etch recipe with repeated DI water quenches used by our CDMS collaborators at TAMU was shown to minimize the undercut of W by the Al, resulting in successful step coverage by the final W film[12].

It is well known that H$_2$ produced during Al wet etching can reduce etch homogeneity[15]. We believe the rough side-wall texture observed in our devices after continuous or long Al etches are caused primarily by $H_2$ bubbles. The surfactant added to some Al etchants helps mitigate this problem by improving the surface wetting of the Al and reducing surface tension[16]. We have shown that by periodically interrupting the Al etch, and even interspersing the etch with a series of DI rinses effectively dissipates the $H_2$ bubbles and cools the wafer, resulting in vastly reduced sidewall texture. Results of three different processes using the same commercial etchant (KMG 16:1:1:2 NP) but different time profiles are shown in Figure 2. It is clear that the etched Al side walls are significantly smoother when the etch is performed as a series of dips rather than as an extended single dip.

## *C. Growth of Sputtered W on Etched Al Surfaces*

We have found that the etched Al sidewalls of our devices must be relatively free of rough texture for 40 nm-thick sputtered W film to attach. Smoothest sidewalls are achieved with short (10-30 sec. depending on etchant) dips in etchant interspersed with DI water quenches. This routine is overly burdensome for ~kg scale CDMS detector processing so a 'finishing etch' method was developed. Sufficiently smooth Al sidewalls are achievable by doing a continuous etch to get through the full Al thickness, followed by three short finishing etches (10 sec. etchant dips/20 sec. DI dips). A comparison of sputtered W film growth on Al features etched in four different ways is shown in Figure 3. When using the long initial etch combined with a finishing etch, line width uniformity was compromised slightly by the formation of cavities 100-400 nm in diameter, but the effect was sufficiently minor to not require an all-short-dip etch process. Another successful fabrication recipe for W/Al QETs was developed by our TAMU collaborators who fabricate devices using a planar sputtering tool with substantially larger targets[12].

## III. Film Step Coverage Model

For CDMS compatibility, our AJA ATC-2200 sputtering system was designed to accommodate four 100 mm diameter x 33.3 mm-thick ~1.4 kg (each) substrates at a time and provide *in vacuo* substrate flipping for depositions on both sides. The wafer carrier rotates about its center at ~15.5 rev/min during depositions and is suspended ~ 8" above the 4" diameter targets (racetrack ~3.5" in diameter), which sputter upwards. A schematic diagram of the sputtering geometry is shown in Figure 4. Wafer flats are always oriented towards the outside of the carrier for diagnostic purposes.

To better understand the origin of irregularities in device performance, a model of the step coverage across the 10" substrate carrier was developed using thickness data obtained from an Al deposition on Si test wafers performed without platter rotation. The thickness data fit a 2-D Gaussian fairly well, but a 2-D King fit (flattened Gaussian) with flattening factor $1/\gamma = 0.1$ was more accurate for film locations farthest from the sputtering target. Mathematically rotating this distribution around the carrier rotation axis yielded a predicted thickness distribution across the full wafer carrier. The results were consistent with measured thin film data from depositions made with carrier rotation, although the model would benefit from introducing an ellipsoidal distribution to represent the gun tilt more accurately.

As is evident from the geometry shown in Figure 4, die locations closer to the axis of rotation are more likely to achieve adequate step coverage than devices very near or very far from the gun, because the angle of incident flux is favorable and the deposition rate is still high. In order to model the deposition on a step, the angle between the carrier face and the gun central axis was calculated and the sine of this angle was multiplied by the 2-D King non-rotating platter thickness distribution. Integrating around a circle of fixed radius from the carrier center then gave a predicted total film thickness on a step edge. By simulating various orientations of the step on the wafer we found that the step edges parallel-to and facing the flat should have the worst step coverage, and edges aligned perpendicular to the flat should have the best step coverage. Orientation dependent step coverage was confirmed in subsequent SEM imaging studies. Figure 5 shows our simulation for step coverage on a step parallel to and facing the flat where simulations show that step coverage goes to zero. Figure 6 shows the step coverage for 600 nm Al devices near the flat compared to near the center of the platter. Limited step coverage in areas near the flat suggests that our model would improve by taking into account the scattering of target atoms by the Ar gas.

## IV. Discussion

## A. W film structure

The zone structure model developed by Thorton[17,18] is useful in understanding the growth of W films on the Al steps. This model describes the microstructure of sputtered films as a function of chamber pressure and $T/T_m$ where $T$ is the local temperature at the deposition site and $T_m$ is the melting point of the sputtered material. For our depositions, $T/T_m$ is low ($T_m$ for W is 3683 K) and the Ar sputter pressure is 3 mTorr so we can limit ourselves to Zone 1 or Zone T in the Thorton model. In Zone 1, adatom diffusion is limited and sputtered W atoms tend to adhere where they strike. This phenomenon tends to create open boundaries on steps because high points receive more flux than low ones, especially when an oblique component to the flux is present, like in our confocal geometry[17]. Ion bombardment of biased substrates can be used to simulate an increase in $T/T_m$, and suppress Zone 1 structure, promoting step coverage[18]. However, when we varied the DC bias from -100 V to -250 V during W depositions on rough Al surfaces, we saw no improvement in step coverage. The simulated increase in $T/T_m$ was not enough to promote film growth. Instead, we found that the best strategy to achieve W-on-Al step coverage was to smooth out the surface on which the 40 nm-thick W would be deposited by modifying our pre-deposition Al wet etch.

Sputtered W thin films can consist of single crystal $\alpha$-phase (bcc, $T_c = 15 mK$), single crystal $\beta$-phase (A15, $T_c = 1-4K$), or a mix of the two phases. Work has been done to show that W $T_c$ is tunable by varying target power and chamber pressure[19]. The current CDMS detector specification for W-film superconducting $T_c \approx 65 mK$ translates into a need for mixed-phase $\alpha - \beta$ W films. Several studies have been performed by our group and others that relate W film crystalline phase and $T_c$, film stress, resistivity, and other properties. Studies show a rapid change from tensile to compressive film stress for W films deposited at low pressure[20-24]. The compressive stress is generally attributed to atomic peening caused by energetic sputter atoms[20]. Our data for ~40 nm-thick W films are consistent with expectations: we observe compressive film stress of 0.7-2 GPa in W for Ar deposition pressures of 3-5 mTorr. Using X-ray diffraction, Haghiri-Gosnet determined the structure of compressively stressed films to consist entirely of $\alpha$-phase

W[20]. Using TEM, the microstructure of compressive films were observed by Shen and Vink. Shen found that compressively stressed W films are either entirely $\alpha$-phase, or a mixture of $\alpha$- and $\beta$-phase W[21]. Vink found that compressively stressed films consist of mostly $\alpha$-phase W with some $\beta$-phase also present[22]. Our $T_c$ measurements ranging from 60-100 mK (but with sharp individual film transitions of < 1 mK) strongly suggest a mix of $\alpha$ and $\beta$ phases in our compressively stressed films. Measured stress in the final W layer of our devices was typically ~1-2 GPa (compressive) and W film resistivity was $\sim 15 \mu\Omega - cm$.

## B. Quantitative Step Coverage Diagnostics

Comprehensive cryogenic studies of the quality of our W/Al step coverage were performed at 35 mK in a dilution refrigerator. The results of these studies correlated well with SEM diagnostic data. An image of one test device is shown in the inset of figure 7, where two 250 um-wide x 250 um-long x 40 nm-thick W-TESs overlap the edges of a 350 um-long by 250 um-wide x 300 nm-thick sputtered Al film. The entire structure is surrounded by a ring of smaller W/Al QETs used to veto substrate events in diagnostic experiments. In these experiments, devices are exposed to a collimated source of 2.62 keV $K_\alpha$ x-rays. The quality of the step coverage at the W/Al interface can be studied quantitatively with this set up[13].

Data obtained with a well-processed (typical) device along with a device with poor W/Al step coverage with the same dimensions is shown in figure 7. Each data point corresponds to a 2.62keV x-ray striking the central Al film. Event energy is collected using the attached W-TESs at the two ends of the Al film. An x-ray entering the detector closer to one W-TES than the other will result in increased energy collection in that TES. The band of events obtained from a device with good W/Al step coverage (black dots in Figure 7) exhibits ~2x more total energy collection than a device with poor step coverage (magenta x's in Figure 7) due to better coupling between the W and Al films. Even with good step coverage, the full 2.62keV is not recovered. Our results are consistent with known quasiparticle down-conversion processes where energy is lost to the substrate[13]. This method provides valuable insight to the properties of the W/Al interface.

## V. SUMMARY AND CONCLUSIONS

Quasiparticle-trap-assisted electrothermal-feedback transition-edge sensors (QETs) are key to our search for weakly interacting massive particles. In order to optimize sensitivity, W TESs must contain a minimal amount of material. In order to transport more phonon energy to the TES, the thickness of the attached Al fin was increased. Thorton's zone models suggest that smoother substrates require a lower $T/T_m$ to promote continuous film growth. Step coverage by a 40 nm W deposition on 600 nm-thick Al features was achieved by etching the Al with KMG 80:3:15 NP (cycles of 20 sec. etch interspersed with 20 sec. quench). Further investigations will be made to quantify the improved detector performance and studies will be carried out to analyze device performance for dies across the platter[25]. If the thin step coverage near the wafer flat is not thick enough to support quasiparticle energy transport without driving the film normal, we will have to resort to doing single substrate depositions at the center of the platter where step coverage is thicker. The results of these studies will feed back into the fabrication recipe for the next generation of SuperCDMS detectors.

## ACKNOWLEDGMENTS

The authors thank Andrew Jastram, Mark Platt, Rusty Harris, and Rupak Mahapatra for useful discussions. The devices were fabricated in the Stanford Nanofabrication Facility. SEM and AFM work was done at Santa Clara University's Center for Nanostructures. FIB work was performed by Evans Analytical Group. This work was supported by the U.S. Department of Energy (DE-FG02-13ER41918) and by the National Science Foundation (PHY-1102842).

[4] R. Horansky, G. Stiehl, J. Beall, K. D. Irwin, A. Plionis, M. Rabin, J. Ullom, *J. of Appl. Phys.* **107**, 044512 (2010)

[5] P. A. R. Ade, R. W. Aikin, D. Barkats, S. J. Benton, C. A. Bischoff, J. J. Bock, J. A. Brevik, I. Buder, E. Bullock, J. J. Bock, *Phys. Rev. Lett.* **112**, 241101 (2014)

[6] M. R. J. Palosaari, K. M. Kinnunen, J. Julin, M. Laitinen, M. Napari, T. Sajavaara, W. B. Doriese, J. Fowler, C. Reintsema, D. Swetz, et. al., *J. of Low Temp. Phys.* **176**, 285 (2014)

[7] K. D. Irwin, S. W. Nam, B. Cabrera, B. Chugg, and B. A. Young, *Rev. Sci. Instrum.* **66**, 5322 (1995)

[8] D. S. Akerib, J. Alvaro-Dean, M. S. Armel-Funkhouser, M. J. Attisha, L. Baudis, D. A. Bauer, J. Beaty, P. L. Brink, R. Bunker, S. P. Burke, et. al., *Phys. Rev. Lett.* **93**, 211301 (2004)

[9] Z. Ahmed, D. S. Akerib, S. Arrenberg, C. N. Bailey, D. Balakishiyeva, L. Baudis, D. A. Bauer, P. L. Brink, T. Bruch, R. Bunker, et. al., *Science* **327**, 1619 (2010)

[10] R. Agnese, A. J. Anderson, M. Asai, D. Balakishiyeva, R. Basu Thakur, D. A. Bauer, J. Beaty, J. Billard, A. Borgland, M. A. Bowles, et. al., Phys. Rev. Lett. **112**, 241302 (2014)

[11] P. L. Brink, Z. Ahmed, D. S. Akerib, C. N. Bailey, D. Balakishiyeva, D. A. Bauer, J. Beaty, R. Bunker, B. Cabrera, D. O. Caldwell, et. al., *AIP Conference Proceedings,* **1185**, 655 (2009)

[12] A. Jastram, R. Harris, R. Mahapatra, J. Phillips, M. Platt, K. Prased, J. Sander, and S. Upadhyayula, *See: http://arxiv.org/abs/1408.0295*

[13] J. J. Yen, B. Shank, B. A. Young, B. Cabrera, P. L. Brink, M. Cherry, J. M. Kreikebaum, R. Moffatt, P. Redl, A. Tomada, and E. C. Tortorici *J. Appl. Phys., submitted for publication. See: http://arxiv.org/abs/1406.7308*

[14] K. R. Williams and R. S. Muller, *J. Microelectromech. Syst.* **5**, 256 (1996).

[15] "Aluminum Etching", Microchemicals GmbH , 7 November 2013, See: http://www.microchemicals.com/downloads/application_notes.html
10

| Etchant | $H_3PO_4$ | $CH_3COOH$ | $HNO_3$ | Surfactant | Etch Rate (Å/s) | Etching interval (DI) |
|---|---|---|---|---|---|---|
| Cyantek Al-11 | 72% | 3% | 3% | No | 8 @ 25º C | 30s (30s) |
| KMG 16:1:1:2 NP | 70-80% | 1-5% | 1-5% | Yes | 9 @ 23º C | 10s (20s) |
| KMG 80:3:15 NP | 60-80% | 5-15% | 1-5% | Yes | 10 @ 24º C | 20s (20s) |

Table I. Properties of Premixed Al Etchants.

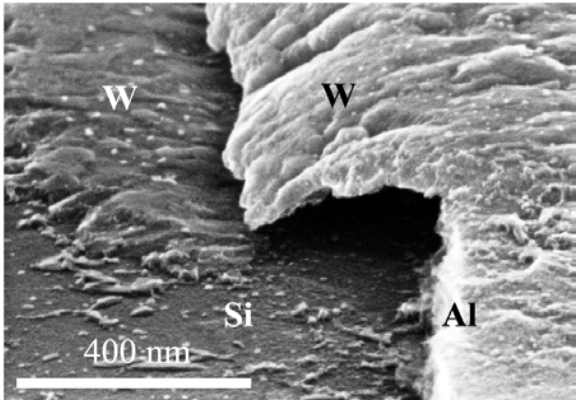

Figure 1. W overhang of Al underetch in a QET after second W deposition. The horizontal W overhang produced during the first etch sequence gets bent towards the wafer's surface during the second W deposition.

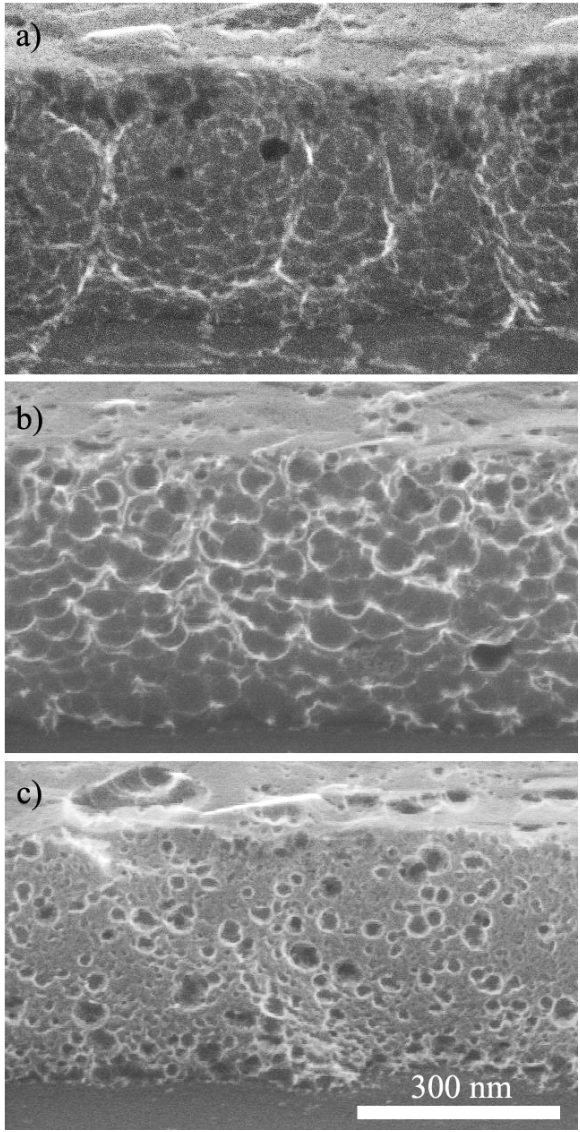

Figure 2. SEM images of sputtered 300 nm-thick Al films etched with various time sequences using KMG 16:1:1:2 NP Al etchant at room temperature. The Al sidewall is smoother when the etch is performed as a series of dips interspersed with DI water quenches rather than extended dips without DI water quenches. (a) An uninterrupted 300 sec. Al etch removed the 300 nm-thick Al film, but left behind rough sidewalls with pronounced cavities ~200-300 nm in diameter. (b) Results of a series of 17 etch/DI quench dips (30 sec. / 30 sec.). The large cavities seen in (a) are no longer present. Rough sidewall features remain. (c) Results from a series of 59 etch/DI quench dips (10 sec. / 20 sec.). Here sidewall roughness is significantly reduced. Cavities ~10 nm in diameter remain.

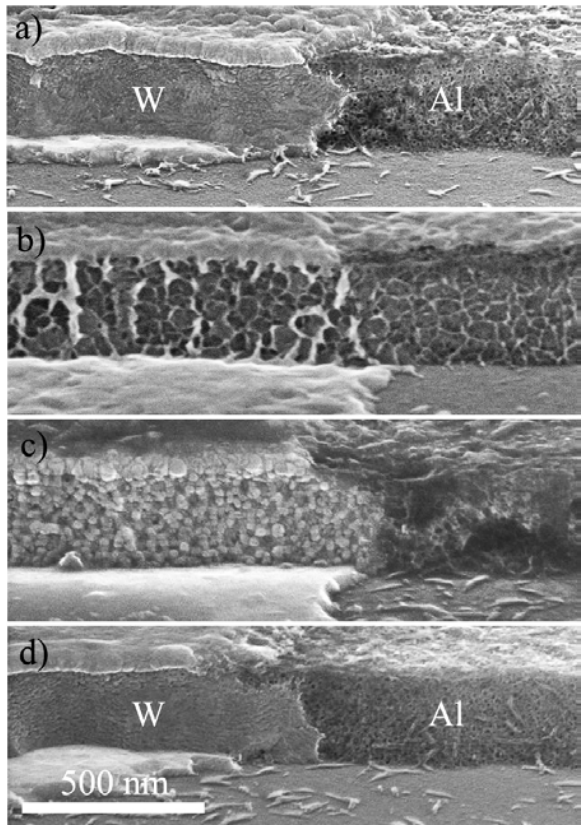

Figure 3. A comparison showing how well 40 nm-thick sputtered W films conform to 300 nm-thick Al films, patterned using four different Al etch sequences. The W deposition parameters were held constant in all cases. (a) The W is observed to adhere uniformly to the Al film etched with (now unavailable) Cyantek AL-11 using 12 etch/ DI quench dips (30 sec. / 30 sec.). We believe this is due to the smooth Al sidewalls produced using this etch sequence. (b) W does not adhere uniformly to Al film etched in Cyantek AL-11 using one 300 sec. dip followed by a DI quench. The rough sidewall surface inhibits W film growth. (c) Al film first etched in KMG 80:3:15 NP for 300 sec. followed by three etch/DI quench dips (10 sec. / 20 sec.). The three short etches were used to smooth out the Al sidewall surface to promote continuous W film growth in this region. The resulting W film was found to be continuous but highly granular. (d) Al film etched with KMG 80:3:15 NP using 25 etch/DI quench dips (20 sec. / 20 sec.). This etch recipe provides smooth Al sidewalls which promote uniform W film growth.

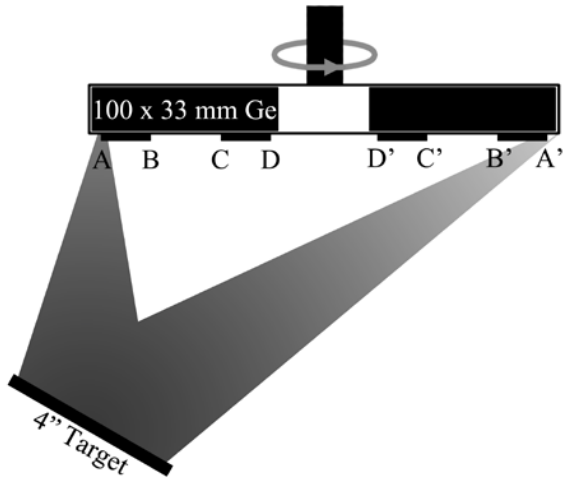

Figure 4. Schematic of deposition geometry where previously patterned features across a 10" diameter substrate platter receive a subsequent deposition. Features at locations A, B, C, and D are rotated during deposition around a central axis (to locations A', B', C', and D'). In this figure, the sidewall of feature A receives predominantly vertical flux whereas A' is in shadow. This results in relatively poor step coverage at the edge of the platter compared to the center.

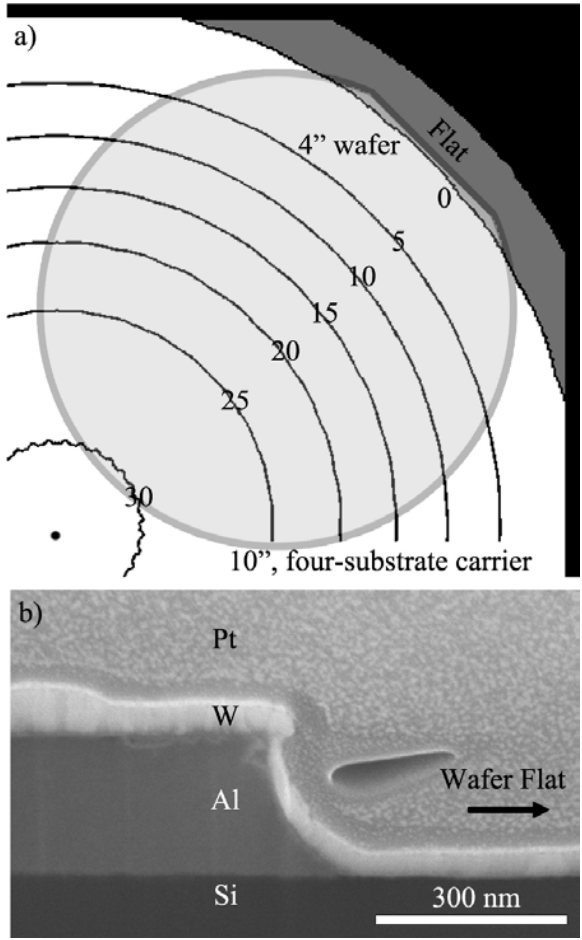

Figure 5. (a) Step coverage simulation (arbitrary units) for features facing the flat in a 10" diameter, four-substrate carrier using a King fit to model deposited film thickness. The small wiggles in the contours is a numerical artifact. (b) FIB image of a conformal coating of a 300 nm-thick Al film with a conformal nominally 40 nm-thick W film at the center of the wafer shown in (a). The feature in (b) faces the wafer flat. The 40 nm-thick W film shows a minimum thickness of 22 nm at this location. (Pt was used to protect the features during FIB imaging.)

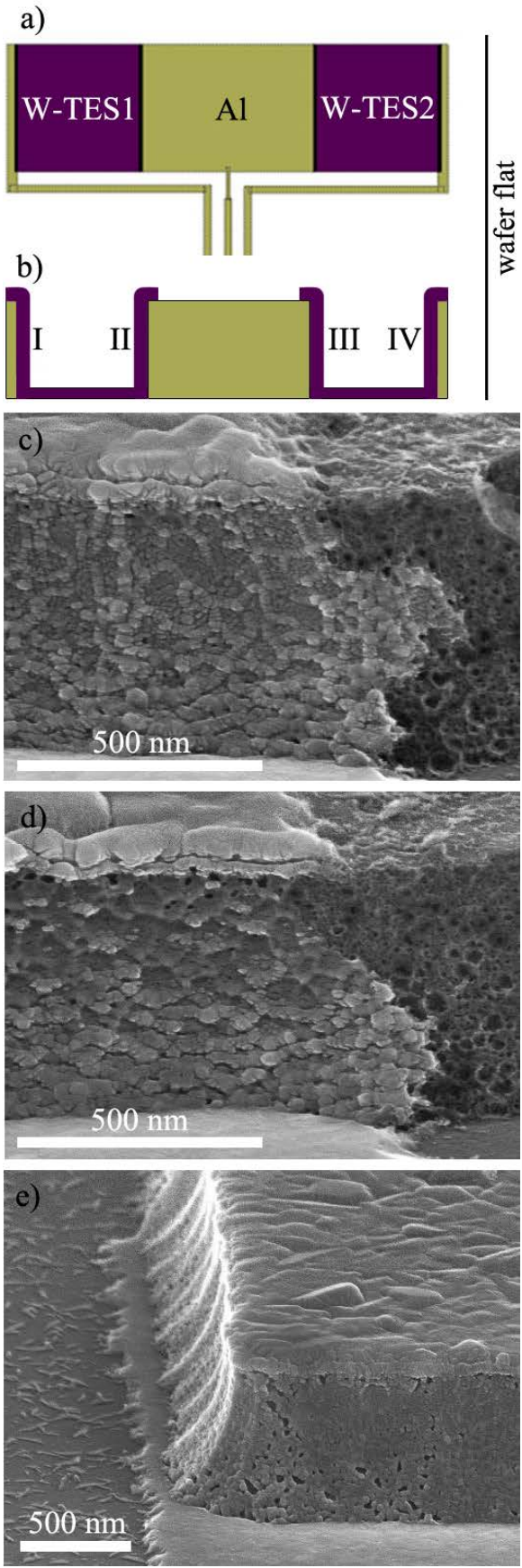

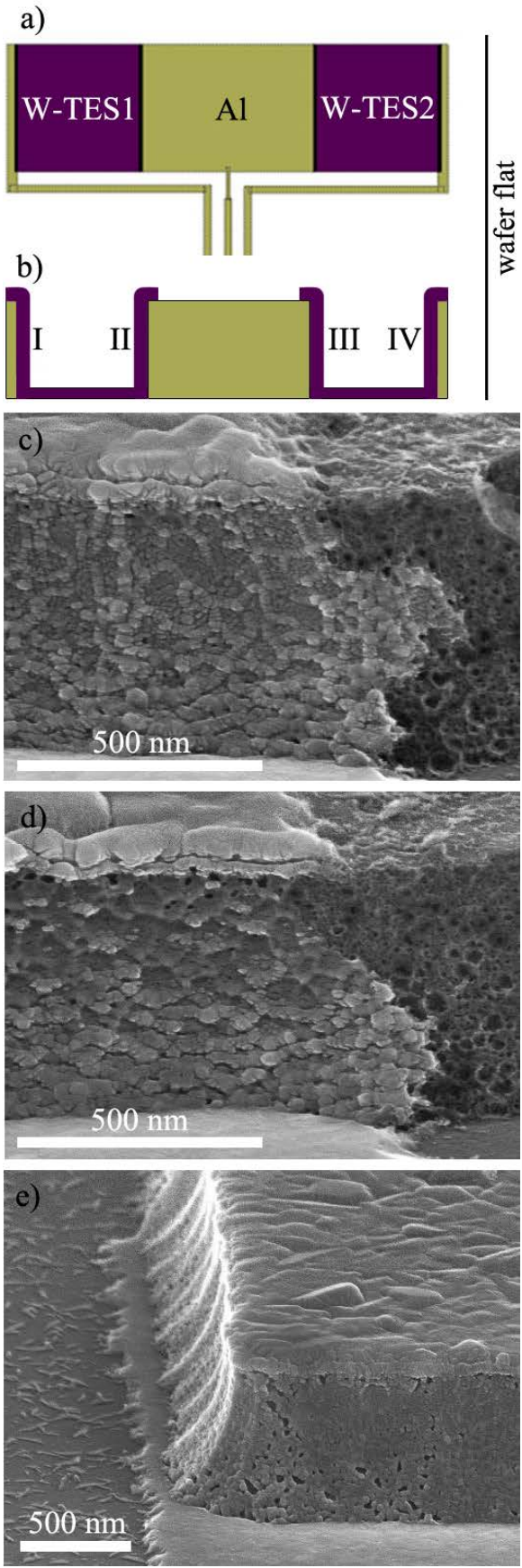

Figure 6. (Color online) QET device where a 40 nm-thick W film achieves continuous step coverage over patterned 600 nm-thick Al features. In our confocal sputtering system, step coverage quality depends on the step's orientation and distance from the center of the wafer carrier. Al films were etched in 26º C KMG 80:3:15 NP with 29 etch/ DI quench dips (20 sec. / 20 sec.). (a) Schematic of QET device where two W-TESs are attached to a central Al phonon collection film. The overlap regions are parallel to the flat of the wafer. (b) Cross-sectional schematic of QET device with film thickness greatly exaggerated. Each QET has four W/Al step coverage regions (I-IV). For a device located in the sputtering chamber 4.5" from the center of the substrate platter, an etched 600 nm-thick Al sidewall in orientation II or IV (c) experiences better step coverage than a sidewall in orientation I or III (d) for 40 nm-thick W, agreeing with simulations. (e) For a device 1.5" from the center of the platter, the step coverage for a sidewall in orientation I or III is improved, also agreeing with simulations.

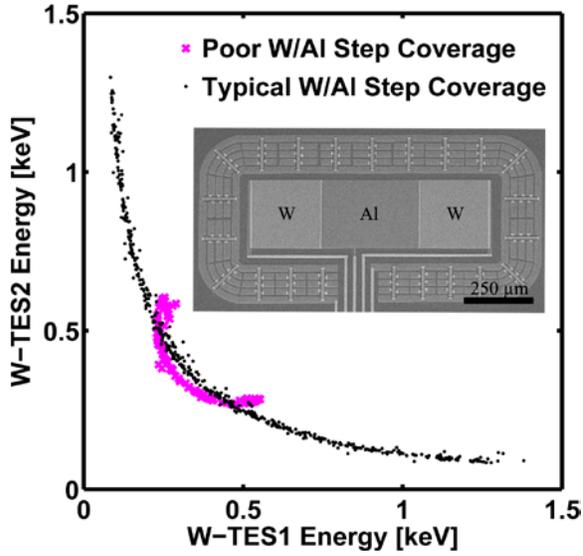

Figure 7. (color online) Results of diagnostic 2.62keV x-ray fluorescence experiments[13] performed at 35 mK to quantitatively evaluate the quality of W/Al interfaces. A SEM of a typical test device is shown (inset). A device with good W/Al interfaces produces a band of events such as that shown in black. A device with poor filamentary W/Al step coverage yields a band of events such as that shown in magenta. More energy is collected from a device with uniform W step coverage.